\documentstyle[12pt]{article}

\advance \textheight by \topskip
\begin{document}

\title{Symmetries and classical quantization}
\author{Khazret S. Nirov${}^a${}\thanks{Alexander von Humboldt fellow;
on leave from the {\it Institute for Nuclear Research, Moscow, 
Russia}}${\,\,}$\thanks{E-mail: nirov@physik.tu-muenchen.de} $\,$
and Mikhail S. Plyushchay${}^{b,c}${}\thanks{E--mail: plyushchay@mx.ihep.su}\\
{\small \it ${}^a{}$Institut f\"ur Theoretische Physik T30,
Physik Department,}\\
{\small \it Thechnische Universit\"at M\"unchen, D-85747 Garching, Germany}\\
{\small \it ${}^b{}$Departamento de Fisica --- ICE, 
Universidade Federal de Juiz de Fora}\\
{\small \it 36036-330 Juiz de Fora, MG Brazil}\\
{\small \it ${}^{c}{}$Institute for High Energy Physics, 
Protvino, Moscow region, 142284 Russia}}

\date{}

\maketitle

\begin{abstract}
\noindent 
A phenomenon of classical quantization is discussed. 
This is revealed in the class of pseudoclassical gauge 
systems with nonlinear nilpotent constraints containing 
some free parameters. Variation of parameters does not 
change local (gauge) and discrete symmetries of the 
corresponding systems, but there are some special 
discrete values of them which give rise to the
maximal global symmetries at the classical level.  
Exactly the same values of the parameters are separated 
at the quantum level, where, in particular, they are 
singled out by the requirement of conservation of the 
discrete symmetries. The phenomenon is observed for the 
familiar pseudoclassical model of $3D$ $P,T$-invariant 
massive fermion system and for a new pseudoclassical 
model of $3D$ $P,T$-invariant system of topologically 
massive U(1) gauge fields.
\end{abstract}

\section{Introduction}
The quantization of parameters takes place in many physically
interesting gauge systems, such as Dirac monopole \cite{mono}, 
\cite{rj}, \cite{dhvin}, non-Abelian topologically massive vector 
gauge theory \cite{nonab} and its particle-mechanics generalization
\cite{ht}, various spin particle models \cite{spin}. 
In particular, this property is specific to some
pseudoclassical spin models \cite{cpv,gps,cp} belonging to the
class of gauge systems with nonlinear nilpotent
constraints \cite{cpv}--\cite{pr}.  The nature of the
quantization phenomenon is hidden in a nontrivial topology of
configuration or phase spaces of the corresponding systems.

Here the following interesting question can be formulated.  If
some system has a quantized parameter, whether its special discrete
values may reveal themselves in some way just at the classical
level?

In this letter we shall discuss exactly such a phenomenon, which
may be called the classical quantization.  We shall
show that the requirement of maximality of classical global
(rigid) symmetry can separate some special discrete values of
corresponding parameters.  These values  turn out to be exactly
the same as those singled out by the quantization procedure.  
Namely, we shall consider a class of pseudoclassical gauge 
systems with quadratic  in Grassmann variables nilpotent 
constraints of the form
$(\alpha_{ik}+\nu\beta_{ik}) \xi_i\xi_k\approx 0$, 
where $\nu$ is a $c$-valued parameter and
$\alpha_{ik}$, $\beta_{ik}$ are functions of even variables.
While varying the values of the parameter $\nu$, we do not change 
local continuous and discrete symmetries of the corresponding
classical system, but may drastically change its 
continuous global symmetries. As a result, there are some 
special discrete values of the parameter, $\nu=\nu_q$, for 
which a classical system has a maximal global symmetry, whose 
set of generators includes the integrals of motion existing only 
at $\nu=\nu_q$.  Exactly the same values of the parameter reveal
themselves at the quantum level too.  But quantum mechanically
they are singled out not only by requiring the continuous global 
symmetry to be maximal.  It turns out that only for these
special values $\nu=\nu_q$ the discrete symmetries of the
corresponding classical system are conserved in the quantum
case.  So, the classical quantization  phenomenon implies also
some non-anticipated hidden relationships between continuous and
discrete symmetries.

The paper is organized as follows.  We start with consideration
of the simplest toy model revealing the described phenomenon.
Then we  investigate in this context the pseudoclassical model 
\cite{cpv,gps} of
$3D$  $P,T$-invariant massive fermion system \cite{pt} and consider 
a new pseudoclassical model for $3D$ $P,T$-invariant system of
topologically massive U(1) gauge fields \cite{pt}--\cite{sie},
\cite{nonab}.  We conclude the paper with some remarks.

\section{Toy model}

Let us consider the following pseudoclassical
model given by the Lagrangian
\begin{equation}
L=v\chi+\frac{i}{2}\xi_a\dot{\xi}_a+
\frac{i}{2}\theta_a\dot{\theta}_a,
\label{lag1}
\end{equation}
where $\xi_a$, $\theta_a$, $a=1,2$, are Grassmann
variables, $v$ is an even Lagrange multiplier, and 
$\chi$ is a nilpotent nonlinear function,
\begin{equation}
\chi=-i(\xi_1\xi_2+\nu\theta_1\theta_2),
\label{con}
\end{equation}
$\chi^3=0$, containing 
a real $c$-number parameter $\nu$.
When introducing the notation
$\xi^\pm=\frac{1}{\sqrt{2}}(\xi_1\pm i\xi_2)$,
$\theta^\pm=\frac{1}{\sqrt{2}}(\theta_1\pm i\theta_2)$,
one can present $\chi$ in the form  
$\chi=\xi^+\xi^-+\nu\theta^+\theta^-$.

Lagrangian (\ref{lag1}) leads to  the nontrivial Dirac brackets 
$
\{\xi^+,\xi^-\}=-i,
$
$
\{\theta^+,\theta^-\}=-i
$
and $\{v,p_v\}=1$, 
and generates the first class primary, $p_v\approx 0$, 
and secondary, $\chi\approx 0$, constraints \cite{sun}.
The total Hamiltonian is a linear combination
of the constraints, $H=v\chi+up_v$, with
$u=u(\tau)$ being an arbitrary function of 
the evolution parameter. 
The phase space variables satisfy the following
Hamiltonian equations of motion:
$\dot{p_v}=0$, $\dot{v}=u$,
$\dot{\xi}{}^\pm=\pm iv\xi^\pm$,
$\dot{\theta}{}^\pm=\pm i\nu v\theta^\pm$.
Their general solutions are
$p_v(\tau)=p_v(0)$,
$v(\tau)=v(0)+\int_0^\tau u(\tau')d\tau'$,
$\xi^\pm(\tau)=e^{\pm i\omega(\tau)}\xi^\pm(0),$
$\theta^\pm(\tau)=e^{\pm i\nu\omega(\tau)}
\theta^\pm(0)$,
where $\omega(\tau)=\int_0^\tau v(\tau')d\tau'$.
One finds the obvious integrals of motion,
$
{\cal N}_\xi=\xi^+\xi^-,
$
$
{\cal N}_\theta=\theta^+\theta^-.
$
Therefore, our nilpotent constraint is their linear 
combination, $\chi={\cal N}_\xi+\nu{\cal N}_\theta$.
The case of $\nu=0$ is degenerate: in this case
variables $\theta^\pm$ have no dynamics,
$\theta^\pm(\tau)=\theta^\pm(0)$,
being trivial integrals of motion.
Now let us put the question:
are there other values of the parameter
$\nu$ which would be special from the point of view
of dynamics?
Using the explicit solution to the equations of motion,
one finds that such special values are $\nu=\pm 1$.
Only in these cases there are two additional
integrals of motion,
\begin{equation}
{\cal T}^+_+=\xi^+\theta^-,
\quad
{\cal T}^-_+=\theta^+\xi^-
=({\cal T}^+_+)^*
\label{adin}
\end{equation}
for $\nu=+1$, and 
${\cal T}^+_-=\xi^+\theta^+$, 
${\cal T}^-_-=\theta^-\xi^-=({\cal T}^+_-)^*$ for $\nu=-1$.
Via the trivial change of the variables, $\theta^\pm\rightarrow
\theta^\mp$, the second case can be reduced to the first one.
Note that when 
$\nu\neq 1$, 
one can construct the integrals
${\cal T}_\nu^+=\xi^+(\tau)\theta^-(\tau)e^{i(\nu-1)\omega(\tau)}$,
${\cal T}_\nu^-=({\cal T}_\nu^+)^*$,
which are {\it nonlocal in the evolution parameter $\tau$
functions} due to the presence
of the factor $e^{i(\nu-1)\omega(\tau)}$.
They become local integrals (\ref{adin}) at $\nu=1$.
In the same way one can construct
nonlocal integrals which turn into 
the local integrals ${\cal T}^+_-$,
${\cal T}^-_-$ at $\nu = -1$.
Thus, we conclude that there are special
values of $c$-number parameter $\nu$,
$\nu=\pm 1$, at which the system has additional
nontrivial local integrals of motion and so, these cases
can be singled out by requiring the maximal
continuous global symmetry in the system.
This continuous global symmetry is generated by
the integrals 
${\cal N}_\xi,$ ${\cal N}_\theta$ 
and the additional integrals
${\cal T}^\pm_+$ or ${\cal T}^\pm_-$.

To conclude the discussion of the classical theory,
one notes that the system (\ref{lag1})
is invariant also with respect to the discrete
transformation 
\begin{equation}
D: \xi_a\rightarrow (\xi_1,-\xi_2),\quad
D: \theta_a\rightarrow (\theta_1,-\theta_2),
\quad
D: v\rightarrow -v,
\label{d}
\end{equation}
taking place for the arbitrary value of the parameter $\nu$.
As we shall see, the quantum analog of this discrete
symmetry plays very important role.

In correspondence with classical brackets, 
at the quantum level we have 
$[\widehat{\xi}_a,\widehat{\xi}_b]_+=\delta_{ab}$,
$[\widehat{\theta}_a,\widehat{\theta}_b]_+=\delta_{ab}$,
$[\widehat{\xi}_a,\widehat{\theta}_b]_+=0$.  One can realize
$\widehat{\xi}_a$ and $\widehat{\theta}_a$ as the operators
$\widehat{\xi}_a=\frac{1}{\sqrt{2}}\sigma_a\otimes \sigma_3$, 
$\widehat{\theta}_a=\frac{1}{\sqrt{2}}
1\otimes \sigma_a$
acting on the space of functions 
$\Psi=\psi_1\otimes\psi_2$, where $\psi_{1,2}=\psi_{1,2}(v)$
are two-component functions.
Then the quantum analog of the constraint $p_v\approx 0$,
$\widehat{p}_v\Psi=0$ with $\widehat{p}_v=-i\partial/\partial v$,
means that the function
$\Psi$ does not depend on $v$.
Finally, the physical subspace of the system
is singled out by the quantum analog of the 
second constraint,
$\widehat{\chi}\Psi=0$.
There is an operator-ordering ambiguity under construction
of the quantum analog of the nonlinear nilpotent constraint, 
and in general case we have 
\begin{equation}
\widehat{\chi}=\sigma_3\otimes 1+\nu \cdot 1\otimes\sigma_3
+\alpha
\label{ca}
\end{equation}
with the constant $\alpha$ (of order $\hbar$)
characterizing the deviation of the ordering 
from that one corresponding to the classical
ordering in (\ref{con}), i.e. $\alpha=0$
corresponds to the antisymmetrized ordering
of operators $\widehat{\xi}{}^+$, $\widehat{\xi}{}^-$
and $\widehat{\theta}{}^+$, $\widehat{\theta}{}^-$ 
(see ref. \cite{cpv} for more detailed discussion of this point).
In the case of antisymmetrized ordering ($\alpha=0$),
we find that the quantum constraint $\widehat{\chi}\Psi=0$
has nontrivial solutions iff $|\nu|=1$. 
The corresponding physical states in transposed form
are given by
$\Psi_{+,1}^t=(1,0)\otimes(0,1)$ and $\Psi_{+,2}^t=
(0,1)\otimes(1,0)$ for $\nu=+1$,
and $\Psi_{-,1}^t=(1,0)\otimes(1,0)$,
$\Psi_{-,2}^t=(0,1)\otimes(0,1)$ for $\nu=-1$.
Therefore, we see that exactly the same values of the 
parameter $\nu$ turn out to be special from the 
quantum point of view. This is, of course,
not an accidental fact.
Indeed, e.g., in the case 
$\nu=+1$ 
the physical states 
$\Psi_{+,1}$, $\Psi_{+,2}$ 
are the eigenstates of the operators 
$\widehat{\cal N}_\xi = \frac{1}{2}(\sigma_3+1)\otimes 1$, 
$\widehat{\cal N}_\theta = 1\otimes \frac{1}{2}(\sigma_3+1)$,
whereas they are transformed mutually by the operators 
$\widehat{\cal T}^+_+=-\sigma_+\otimes \sigma_-$ 
and 
$\widehat{\cal T}^-_+=-\sigma_-\otimes\sigma_+$.
Therefore, the physical operators 
$\widehat{\cal N}_\xi$, 
$\widehat{\cal N}_\theta$,  
$\widehat{\cal T}^+_+$ and 
$\widehat{\cal T}^-_+$ can be
considered as generators of the corresponding global 
(i.e. constant in $\tau$) symmetry.

At the classical level we had another, dynamically degenerated
special case corresponding to $\nu=0$.  
So, let us put $\nu=0$ and consider
the quantum constraint $\widehat{\chi}$ in the most general form
(\ref{ca}) containing ordering parameter $\alpha$. Then one can
find that the quantum condition $\widehat{\chi}\Psi=0$  has
nontrivial solutions only either at $\alpha=1$ that corresponds to the
normal ordering of the operators 
$\widehat{\xi}{}^+$, $\widehat{\xi}{}^-$
under construction of the quantum analog of our basic nilpotent
constraint, $\widehat{\chi}=\widehat{\xi}{}^+\widehat{\xi}{}^-$,
or at $\alpha=-1$ in the case of antinormal
ordering. In these cases, in correspondence with classical
picture, the operators $\widehat{\theta}{}^\pm$ will be physical
operators additional to the physical operator 
$\widehat{\cal N}_\xi$.

Let us analyze the general quantum case given by arbitrary
values of both parameters $\nu$ and $\alpha$.  One finds
that if $|\nu|\neq0, 1$, the quantum constraint
$\widehat{\chi}\Psi=0$ has nontrivial solution under appropriate
choice of the parameter $\alpha$ (in this case the corresponding
value of this parameter is different from $\pm 1$ and 0), 
however there is {\it only one} corresponding physical state
annihilated by $\widehat{\chi}$.  Therefore, the same values of 
the parameter $\nu$ ($\nu=\pm 1$, $0$) turn out to be special also
from the point of view of the quantum theory if we require the
maximal number of solutions of the quantum constraint 
$\widehat{\chi}\Psi=0$, so that the maximal (global) symmetry
would be realized on the corresponding physical states.

Let us recall that we had also the classical discrete symmetry 
(\ref{d}). What does happen with it upon quantization?
The corresponding unitary operator generating this
discrete symmetry is $U_D=R_v \sigma_2\otimes\sigma_1$,
where the reflection operator $R_v$ is given 
by the relations $R_v^2=1$, $R_v v=-vR_v$.
Operator $U_D$ transforms the state
$\Psi_{+,1}$ into $\Psi_{+,2}$ 
and $\Psi_{-,1}$ into $\Psi_{-,2}$,
and vice versa. 
Therefore, in accordance with the preceding  discussion,
the discrete symmetry (\ref{d}) survives at the quantum
level only when the parameter takes the special
values $\nu=\pm 1$ (and when, therefore, the 
ordering parameter $\alpha$ is zero).

Further we shall see that in two concrete physical models of
$P,T$-invariant $3D$ systems the special values of the
corresponding parameters can also be separated at the quantum
level {\it by requiring the conservation of the discrete
symmetries taking place at the classical level}.  Exactly the
same special discrete values will be singled out classically
and quantum mechanically by requiring the maximality of the
global symmetries, as it just happened in the toy model.

\section{$3D$ massive double fermion system}

Let us consider first the pseudoclassical physical model of $3D$
$P,T$-invariant massive fermion system revealing the structure
similar to that of the toy model.  Such a
model is given by the Lagrangian \cite{cpv,gps}
\begin{equation}
L=\frac{1}{2e}(\dot{x}_\mu-iv\epsilon_{\mu\nu\lambda}
\xi^\nu\xi^\lambda)^2-\frac{e}{2}m^2
+2i\nu mv\theta_1\theta_2
-\frac{i}{2}\xi_\mu\dot{\xi}{}^\mu
+\frac{i}{2}\theta_a\dot{\theta}_a,
\label{fer}
\end{equation}
where $\theta_a$, 
$a=1,2,$ are Grassmann scalar variables,
$\xi_\mu$, $\mu=0,1,2,$ is Grassmann Lorentz vector,
$e$ and $v$ are even Lagrange multipliers, 
$x_\mu$ are space-time 
coordinates of the particle, $m$ is a mass parameter,
$\nu$ is the $c$-number parameter,
and we use the metric $\eta_{\mu\nu}=diag (-,+,+,)$
and the totally antisymmetric tensor
$\epsilon_{\mu\nu\lambda}$, $\epsilon^{012}=1$.
Lagrangian (\ref{fer}) is invariant with respect 
to the discrete $P$ and $T$ transformations \cite{gps}
(cp. with ref. \cite{hht}, where the time-reversal symmetry 
of the relativistic spinning particle is analyzed),
$
P: X_\mu\rightarrow (X_0,-X_1, X_2),
$
$
T: X_\mu\rightarrow (-X_0,X_1,X_2),
$
where $X_\mu=x_\mu,\xi_\mu$, and
$P:,T: (e,v)\rightarrow (e,-v)$
$P:\theta_a\rightarrow (\theta_1,-\theta_2)$,
$T:\theta_a\rightarrow(-\theta_1,\theta_2)$.
It is necessary to stress that the invariance 
with respect to these discrete transformations is valid
{\it classically} for any value of the parameter $\nu$.

The Hamiltonian description of the system is given by 
nontrivial brackets $\{\theta_a,\theta_b\}=-i\delta_{ab}$,
$\{\xi_\mu,\xi_\nu\}=i\eta_{\mu\nu}$,
$\{x_\mu,p_\nu\}=\eta_{\mu\nu}$,
$\{e,p_e\}=1$, $\{v,p_v\}=1$,
by the set of the first class primary,
$p_e\approx 0$, $p_v\approx 0$,
and secondary,
\[
\phi=\frac{1}{2}(p^2+m^2)\approx0,\quad
\chi=i(\epsilon_{\mu\nu\lambda}
p^\mu \xi^\nu \xi^\lambda
-2\nu m\theta_1\theta_2)\approx 0,
\]
constraints. The total Hamiltonian
$H=e\phi+v\chi+u_1p_e+u_2p_v$ contains arbitrary
functions $u_{1,2}=u_{1,2}(\tau)$ \cite{sun} and
generates the following equations of motion:
$\dot{p}_\mu=\dot{p}_e=\dot{p}_v=0$,
$\dot{x}_\mu=ep_\mu+i\epsilon_{\mu\nu\lambda}
\xi^\nu\xi^\lambda$,
$\dot{\xi}_\mu=v\epsilon_{\mu\nu\lambda}p^\nu\xi^\lambda$,
$\dot{\theta}^\pm=\pm i\nu mv\theta^\pm$,
$\dot{e}=u_1$, $\dot{v}=u_2$.
One can immediately identify the essential integrals
of motion: the energy-momentum vector 
$p_\mu$, the total angular momentum vector
$J_\mu=-\epsilon_{\mu\nu\lambda}x^\nu
p^\lambda-\frac{i}{2}\epsilon_{\mu\nu\lambda}
\xi^\nu\xi^\lambda$, and 
$\Gamma=p\xi$, ${\cal N}_\theta=\theta^+\theta^-$,
$\Delta=i\epsilon_{\mu\nu\lambda}
p^\mu\xi^\nu\xi^\lambda$.
Therefore, the nilpotent constraint is again a
linear combination of the integrals of motion,
$\chi=\Delta+2\nu m {\cal N}_\theta$.
The scalar Grassmann variables $\theta^\pm$
have the dynamics of the type we had
in the toy model,
$
\theta^\pm(\tau)=e^{\pm i\nu\omega(\tau)}\theta^\pm(0),
$
where now $\omega(\tau)=m\int_0^\tau v(\tau')d\tau'$.

Using the mass shell constraint $\phi\approx 0$,
one may introduce the complete oriented triad
$e^{(\alpha)}_{\mu}=e^{(\alpha)}_\mu(p)$, $\alpha=0,1,2$,
$
e^{(0)}_\mu={p_\mu}/{\sqrt{-p^2}},
$
$
e^{(\alpha)}_\mu\eta_{\alpha\beta}e^{(\beta)}_\nu=\eta_{\mu\nu},
$
$
\epsilon_{\mu\nu\lambda}e^{(0)\mu}e^{(i)\nu}e^{(j)\lambda}=
\epsilon^{0ij}.
$
When defining
$\xi^{(\alpha)}=\xi^{\mu}e^{(\alpha)}_\mu$
and
$
\xi^{(\pm)}=\frac{1}{\sqrt{2}}(\xi^{(1)}\pm i\xi^{(2)}),
$
one has
$
\{\xi^{(+)},\xi^{(-)}\}=i,
$
which differs in sign from the brackets for the variables
$\theta^\pm$.  It is necessary to note that the space-like
components of the triad $e^{(i)}_\mu$, $i=1,2$, are not Lorentz
vectors (see, e.g. ref. \cite{spin}), and so, the quantities
$\xi^{(i)}$ as well as $\xi^{(\pm)}$ are not Lorentz scalars.  With
the help of the mass shell constraint, the nilpotent constraint
can be presented in the equivalent form 
$\chi=2m({\cal N}_\xi-\nu{\cal N}_\theta)$, 
where ${\cal N}_\xi=\xi^{(+)}\xi^{(-)}$ is the integral of
motion coinciding up to the sign with the spin, 
${\cal N}_\xi=-J^{(0)}$.
The variables $\xi^{(\pm)}$ have the evolution law analogous to that
in the toy model: $\xi^{(\pm)}(\tau)=e^{\pm
i\omega(\tau)}\xi^{(\pm)}(0)$.
Therefore, exactly as in the case 
of the toy model, we have additional integrals of motion
${\cal T}^\pm_+$ or ${\cal T}^\pm_-$
if and only if $\nu=+1$ or $-1$, 
and they  are given by the same expressions 
(with the change $\xi^\pm\rightarrow \xi^{(\pm)}$)
presented in the toy model.
Of course, the case $\nu=0$ is again degenerated
with variables $\theta^\pm$ being trivial integrals
of motion. As we shall see,  
this case is completely excluded at the quantum level. 
Hence, we conclude that at the classical 
level the discrete values $\nu=\pm 1$ are 
special due to exactly the same reasons we outlined 
in the case of the toy model.

Following the classical brackets, quantum operators 
associated with the odd variables $\xi^\mu$ can be 
realized as 
$\widehat{\xi}^\mu=\frac{1}{\sqrt{2}}\gamma^\mu\otimes \sigma_3$, 
whereas operators 
$\widehat{\theta}_a$ 
can be realized as 
$\widehat{\theta}_a=\frac{1}{\sqrt{2}}1\otimes\sigma_a$.
Here the $\gamma$-matrices satisfy the relation
$\gamma_\mu\gamma_\nu=-\eta_{\mu\nu}
+i\epsilon_{\mu\nu\lambda}\gamma^\lambda$
and can be chosen in the form
$\gamma^0=\sigma_3$, $\gamma^i=i\sigma_i$,
$i=1,2$. 
It is convenient to assume that
the quantum states $\Psi$ 
in transposed form are presented as $\Psi^t=(\psi^t_u,
\psi^t_d)$, and that $\gamma$-matrices act on the spinor
indices of the states $\psi_u$ and $\psi_d$,
whereas $\sigma$-matrices of the second factor
in the operators $\widehat{\xi}_\mu$ and $\widehat{\theta}_a$
act in the space specified by the indices $u$ and $d$.
Quantum constraints
$\widehat{p}_e\Psi=\widehat{p}_v\Psi=0$ with 
$\widehat{p}_e=-i\partial/\partial e$, 
$\widehat{p}_v=-i\partial/\partial v$
mean the independence of physical states from $e$ and $v$,
and the essential quantum conditions are
\begin{equation}
\widehat{\phi}\Psi=0,\quad
\widehat{\chi}\Psi=0.
\label{q1}
\end{equation}
Here the quantum counterpart of the nilpotent constraint
has the form
$
\widehat{\chi}=\widehat{p}_\mu\gamma^\mu\otimes 1
+\nu m(1\otimes \sigma_3+\alpha),
$
where $\alpha=0$ corresponds to the antisymmetrized ordering
of the operators $\widehat{\theta}{}^+$, $\widehat{\theta}{}^-$.
In this case ($\alpha=0$), one can easily check that
two quantum conditions (\ref{q1}) have nontrivial 
solutions iff $\nu=\epsilon=\pm 1$. 
The corresponding  physical states $\psi_u$ and $\psi_d$
form the pair of (2+1)-dimensional Dirac fields,
$(-i\partial_\mu\gamma^\mu+\epsilon m)\psi_u=0$,
$(-i\partial_\mu\gamma^\mu-\epsilon m)\psi_d=0$.
Therefore, in both cases, $\nu=+1$ and $\nu=-1$,
we have a $P,T$-invariant system of two 
fermion fields with opposite spins $+1/2$ and $-1/2$.

Let us assume now that $\alpha\neq 0$.
This corresponds to the operator ordering in 
$\widehat{\chi}$ 
different from the antisymmetrized ordering.
Then we find that at $|\nu|\neq 1$ quantum conditions 
(\ref{q1}) have nontrivial solutions
when the values of $\alpha$ and $\nu$ are related as
$(\nu(\alpha+ 1))^2=1$ or $(\nu(\alpha-1))^2=1$.
In these cases we have as the physical state only 
one of the two, $\psi_u$ or $\psi_d$,
satisfying the corresponding Dirac equation, but not 
the both states. This means that at $|\nu|\neq 1$
only one fermion state of spin $+1/2$ or $-1/2$ 
will be physical, and so, the $P,T$-invariance
of the classical theory will be broken \cite{gps}.
Moreover, we see that $\nu=0$ is completely 
excluded quantum  mechanically since the set of equations
(\ref{q1}) has no nontrivial solutions in
this case.

We conclude that in the case of the pseudoclassical model
given by the Lagrangian (\ref{fer}), the same discrete values 
of the parameter, $\nu=\pm 1$, turn out to be special from the
classical and the quantum points of view.  At $|\nu|=1$ the
quantum analogs of the corresponding additional integrals of
motion together with quantum analogs of integrals ${\cal
N}_\xi$, ${\cal N}_\theta$ generate U(1,1) dynamical symmetry
and $N=3$ supersymmetry, and as it was shown in ref. \cite{gps}, 
the physical states realize irreducible representation
of a nonstandard superextension of the (2+1)-dimensional 
Poincar\'e group.

\section{$P,T$-invariant system of topologically massive
U(1) gauge fields}

Let us consider the pseudoclassical model
of $P,T$-invariant system of topologically massive
U(1) gauge fields.
The novel feature which will be revealed here
is that the corresponding additional integrals of
motion taking place at special values of the corresponding
$c$-number parameter $\nu$ will be of the third
order in Grassmann variables though the corresponding 
nilpotent constraint will again be quadratic.

The model is given by the Lagrangian
\begin{equation}
L=\frac{1}{2e}\left(\dot{x}_\mu-\frac{i}{2}v\epsilon_{\mu\nu\lambda}
\xi^\nu_a\xi^\lambda_a\right)^2-\frac{e}{2}m^2
-i\nu mv\xi_1^\mu\xi_{2\mu}
-\frac{i}{2}{\xi}_{a\mu}\dot{\xi}{}^\mu_a,
\label{vector}
\end{equation}
where now we have two Grassmann Lorentz vectors,
$\xi^\mu_a$, $a=1,2$, instead of one 
vector $\xi_\mu$ and two scalar variables $\theta_a$
from the previous model.
The model is invariant with respect to $P$
and $T$ transformations, under which
the variables $x_\mu$, $e$ and $v$ 
are transformed in the same way as in the previous model,
$\xi_1^\mu$ is transformed as $\xi^\mu$,
while for $\xi_2^\mu$ we have additional 
sign factor in comparison with the transformation of 
$\xi_1^\mu$, 
$P:\xi_2^\mu\rightarrow-(\xi_2^0,-\xi_2^1,\xi^2_2)$,
$T:\xi_2^\mu\rightarrow-(-\xi_2^0,\xi^1_2,\xi^2_2)$,
i.e. we suppose that $\xi^\mu_2$ is a pseudovector.
Again, classically $P$- and $T$-invariance take place 
for arbitrary values of the parameter $\nu$.

Lagrangian (\ref{vector}) generates the constraints
of the same form as we had in the fermion model,
with the only difference in the structure of the nilpotent
constraint. 
Now it is given by 
\begin{equation}
\chi=\frac{i}{2}(\epsilon_{\mu\nu\lambda}
p^\mu\xi_a^\nu\xi_a^\lambda+
2\nu m\xi_1\xi_2)\approx 0.
\label{convec}
\end{equation}
The obvious essential integrals of motion are 
energy-momentum vector $p_\mu$
and the total angular momentum vector
$J_\mu=-\epsilon_{\mu\nu\lambda}x^\nu
p^\lambda-\frac{i}{2}\epsilon_{\mu\nu\lambda}
\xi_a^\nu\xi_a^\lambda$.

It is convenient to introduce the complex 
mutually conjugate vector variables,
$b^\pm_\mu=\frac{1}{\sqrt{2}}(\xi_{1\mu}\pm i\xi_{2\mu})$.
They have the following brackets:
$\{b^+_\mu,b^-_\nu\}=i\eta_{\mu\nu}$,
$\{b^+_\mu,b^+_\nu\}=\{b^-_\mu,b^-_\nu\}=0$.
Equations of motion for these variables have the form
\begin{equation}
\dot{b}{}^\pm_\mu=v(\epsilon_{\mu\nu\lambda}
p^\nu\pm i \nu m \eta_{\mu\lambda})b^{\pm\lambda}.
\label{bpm}
\end{equation}
Using the triad $e^{(\alpha)}_\mu(p)$
and notations $b^{(\alpha)\pm}=b^\pm_\mu e^{(\alpha)\mu}$,
we find the general solution to the equations
of motion for odd variables
\[
b^{(0)\pm}(\tau)=
e^{\pm i\nu \omega(\tau)}b^{(0)\pm}(0),\quad
b^{(i)\pm}(\tau)=e^{\pm i\nu\omega(\tau)}
\left(\cos\omega(\tau)b^{(i)\pm}(0)-\epsilon^{ij}
\sin\omega(\tau)b^{(j)\pm}(0)\right),
\]
where $\epsilon^{ij}=\epsilon^{0ij}$,
$\omega(\tau)=m\int_0^\tau v(\tau')d\tau'$.
{}From here we get the 
quadratic in odd variables integrals of motion, 
$-i\epsilon^{ij}b^{(i)+}b^{(j)-}=J^{(0)}$ 
(spin of the system), and 
${\cal N}_0=b^{(0)+}b^{(0)-}$, ${\cal N}_\bot=
b^{(i)+}b^{(i)-}$.
With the help of the mass shell constraint,
the nilpotent constraint is presented as a linear
combination of the integrals,
$\chi=-m(J^{(0)}+\nu(-{\cal N}_0 +{\cal N}_\bot))$.
Again, we find that the case $\nu=0$ is special:
it gives dynamically trivial variables
$b^{(0)\pm}$,
$b^{(0)\pm}(\tau)=b^{(0)\pm}(0)$,
and as we shall see, this value
of $\nu$ will be excluded at the quantum level.
In order to reveal nontrivial special values of the 
parameter, let us construct the 
following quadratic in odd variables combinations:
$A^\pm=(b^{(2)+}b^{(2)-}-
b^{(1)+}b^{(1)-})\pm i(b^{(2)+}b^{(1)-}+
b^{(1)+}b^{(2)-}).$
They satisfy a simple evolution law:
${A}^\pm(\tau)=e^{\mp 2i\omega(\tau)}A^\pm(0)$.
We immediately conclude that 
when $|\nu|=2$, there are
two additional integrals of motion of the third
order in odd variables. They are
${\cal A}^+_+=A^+b^{(0)+}$,
${\cal A}^-_+=({\cal A}^+_+)^*$ 
for $\nu=+2$ and 
${\cal A}^+_-=A^+b^{(0)-}$,
${\cal A}^-_-=({\cal A}^+_-)^*$ for $\nu=-2$,
and so, this model reveals the classical quantization
of the parameter.

Let us quantize the model.  In correspondence with classical
brackets, the quantum counterparts of the odd variables 
$b^\pm_\mu$ have to form the algebra of fermionic 
creation-annihilation operators,
$[\widehat{b}{}^+_\mu,\widehat{b}{}^-_\nu]_{{}_+}=-\eta_{\mu\nu}$,
$[\widehat{b}{}^+_\mu,\widehat{b}{}^+_\nu]_{{}_+}=
[\widehat{b}{}^-_\mu,\widehat{b}{}^-_\nu]_{{}_+}=0$.
After taking into account the quantum analogs of the constraints
$p_e\approx 0$, $p_v\approx 0$,
two remaining quantum constraints are to annihilate
the states of the general form
\[
\Psi(x)=\left(f(x)+{\cal F}^{\mu}_-(x)\widehat{b}{}^+_\mu
+\epsilon_{\mu\nu\lambda}{\cal F}{}^\mu_+(x)
\widehat{b}{}^{+\nu}\widehat{b}{}^{+\lambda}
+\tilde{f}(x)\epsilon_{\mu\nu\lambda}\widehat{b}{}^{+\mu}
\widehat{b}{}^{+\nu}\widehat{b}{}^{+\lambda}\right)
|0\rangle,
\]
with the vacuum state $|0\rangle$
defined by $\widehat{b}{}^{-}_\mu|0\rangle=0$. 
The quantum $P$ and $T$ transformations are given here as
$
P,T:\Psi(x)\rightarrow \Psi'(x'_{P,T})=U_{P,T}\Psi(x),
$
$
x'{}^\mu_P=(x^0,-x^1,x^2),
$
$
x'{}^\mu_T=(-x^0,x^1,x^2),
$
where
$U_P=B^0_+ B^1_- B^2_+$,
$U_T=B^0_- B^1_+ B^2_+$,
$B^\mu_\pm=\widehat{b}{}^{+\mu}\pm \widehat{b}{}^{-\mu}$.
Operators $U_P$ and $U_T$ are antiunitary,
and in correspondence with classical relations,
they give 
$U_P \widehat{b}{}^\pm_{0,2}U^{-1}_P=\widehat{b}{}^{\mp}_{0,2}$,
$U_P \widehat{b}{}^\pm_{1}U^{-1}_P=-\widehat{b}{}^{\mp}_{1}$,
$U_T \widehat{b}{}^\pm_{1,2}U^{-1}_T=\widehat{b}{}^{\mp}_{1,2}$,
$U_T \widehat{b}{}^\pm_{0}U^{-1}_T=-\widehat{b}{}^{\mp}_{0}$.
One can easily check that while 
acting on the field $\Psi(x)$,
these operators transform scalar
fields $f(x)$ and $\tilde{f}(x)$
one into another, and induce mutual transformation
of the vector fields ${\cal F}_+^\mu(x)$ and 
${\cal F}_-^\mu(x)$.

The physical states are singled out by the quantum 
analogs of remaining two first class constraints,
$(-\partial^2+m^2)\Psi=0$, $\widehat{\chi}\Psi=0$.
Let us fix the same ordering
in the quantum operator $\widehat{\chi}$ 
as in the classical constraint (\ref{convec}). 
This gives $\widehat{\chi}= \epsilon_{\mu\nu\lambda}
\widehat{b}{}^{+\mu}\widehat{b}{}^{-\nu}\partial^\lambda
-\nu m(\widehat{b}{}^{+}_\mu\widehat{b}{}^{-\mu}+3/2)$.
As a consequence of the quantum constraints,
we immediately find that
$f(x)=\tilde{f}(x)=0$,
whereas for the fields 
${\cal F}^\mu_\pm$ 
we have the equations 
\begin{equation}
\left(\epsilon_{\mu\nu\lambda}\partial^\nu
\mp\frac{1}{2}\nu m\eta_{\mu\lambda}\right){\cal F}^\lambda_\pm=0,
\label{lin}
\end{equation}
and $(-\partial^2+m^2){\cal F}^\mu_\pm=0$.
It is interesting to note that
equations (\ref{lin})
contain the tensor operator being the quantum counterpart 
of the classical tensor quantity generating
the classical equations of motion (\ref{bpm})
for $b^\pm_\mu$.
Due to the linear equations (\ref{lin}),
we have also
$\partial_\mu{\cal F}{}^\mu_\pm=0$
and $(-\partial^2+\frac{1}{4}\nu^2m^2){\cal F}^\mu_\pm=0$.
Hence, the quantum constraints are consistent 
(i.e. have nontrivial solutions) if and only if
$|\nu|=2$, i.e. we arrive at the same quantization
condition which we have obtained at the 
classical level. 
Putting $\nu=2\epsilon$, $\epsilon=+$ or $-$,
we get finally that the quantum analog of the nilpotent
constraint gives us the equations
for $P,T$-invariant system of topologically massive
U(1) gauge fields carrying spins $+1$ and $-1$ \cite{pt}.

If we choose another ordering prescription
under construction of the quantum analog of the 
constraint $\chi$, we would have the same
quantum operator but with the constant term $3/2$
changed for $3/2+\alpha$, where constant 
$\alpha$ specifies the ordering.
As a result, under appropriate choice of the parameter $\nu$,
$|\nu|\neq 2,0$, for $\alpha\neq 0, -3/2,+3/2$ we would have, 
as a solution of the quantum constraints, only one field ${\cal
F}_+^\mu$ or ${\cal F}_-^\mu$ satisfying the corresponding
linear differential equation. In this case we would have
violation of the discrete $P,T$-symmetries taking place in the
system at the classical level.

In the cases $\alpha=-3/2$ (or $\nu=0$) or $\alpha=+3/2$, the
physical states would be given by one scalar field $f(x)$ or
$\tilde{f}(x)$, respectively, and in correspondence with the
discussion above, for both these cases we again lose $P$- and
$T$-invariance.

Therefore, the same values of the parameter $\nu$, $\nu=\pm 2$,
which we have separated classically, turn out to be special
quantum mechanically: in these cases the number of physical
states is maximal, and only at $\nu=\pm 2$ we conserve the
$P,T$-symmetries.

One can show \cite{pre} that the quantum counterparts 
of the additional integrals of motion 
${\cal A}^\pm_+$ or ${\cal A}^\pm_-$
give rise to the hidden
U(1,1) symmetry and $N=3$ supersymmetry,
as it takes place for the 
$P,T$-invariant massive fermionic system.

\section{Concluding remarks}

We have revealed a phenomenon of the classical quantization for
the particular class of pseudoclassical systems containing
nilpotent quadratic in Grassmann variables constraints
\cite{cpv}--\cite{pr}.  
The peculiarity generic to this class of constraints is 
that they admit no, even local, gauge conditions \cite{pr}.  
Following the ideas of the present paper it would be 
interesting to investigate other known pseudoclassical spin
particle models \cite{cp}-\cite{pseudo7} belonging to this class. 
It seems very likely that the same phenomenon can also be revealed 
in pseudoclassical systems with higher (than second) order nilpotent 
constraints belonging to the same peculiar class of constraints 
\cite{pr}.

One might try to generalize the class of models we have considered
to the case of systems with infinite number of odd degrees
of freedom. One could this way arrive at some
interesting from the physical point of view spin chain 
systems revealing the phenomenon of classical quantization.

Concluding, it is of interest to investigate in the same context
non-Grassmannian mechanical \cite{spin} and field
\cite{mono,nonab} gauge systems with quantized parameters.

\small
\vskip0.5cm
{\bf Acknowledgement}
\vskip0.3mm

The work of Kh.N. has been supported by the Alexander von Humboldt
Fellowship and by the European Commission TMR programme 
ERBFMRX--CT96--0045 and ERBFMRX--CT96--0090. 
M.P. thanks Prof. H. P. Nilles and TU M\"unchen, where the part of 
this work has been realized, for kind hospitality.


\begin{thebibliography}{**}

\bibitem{mono}
P. A. M. Dirac, {\it Proc. Roy. Soc.} (London)
{\bf A133} (1931) 60

\bibitem{rj}
R. Jackiw, {\it Ann. Phys.} {\bf 129} (1980) 183

\bibitem{dhvin}
E. D'Hoker and L. Vinet, {\it Phys. Lett.} {\bf 137B}
(1984) 72

\bibitem{nonab}
S. Deser, R. Jackiw and S. Templeton,
{\it Phys. Rev. Lett.} {\bf 48} (1982) 975,
{\it Ann. Phys.} {\bf 140} (1982) 372

\bibitem{ht}
P. S. Howe and P. K. Townsend,
{\it Class. Quantum Grav.} {\bf 7} (1990) 1655

\bibitem{spin}
M. S. Plyushchay, {\it Mod. Phys. Lett.}
{\bf A4} (1989) 837;
{\it Phys. Lett.} {\bf B236} (1990) 291,
{\bf B240} (1990) 133, {\bf B248} (1990) 299

\bibitem{cpv}
J. L. Cort\'es, M. S.  Plyushchay and L. Vel\'azquez,
{\it Phys. Lett.} {\bf B306} (1993) 34

\bibitem{gps}
G. Grignani, M. Plyushchay and P. Sodano, 
{\it Nucl. Phys.} {\bf B464} (1996) 189

\bibitem{cp}
J. L. Cort\'es and M.S. Plyushchay,
Preprint hep-th/9602106

\bibitem{pseudo1}
L. Brink, P. Di Vecchia and P. S. Howe,
{\it Nucl. Phys.} {\bf B118} (1977) 76

\bibitem{pseudo2}
V. D. Gershun and V. I. Tkach,
{\it JETP Lett.} {\bf 29} (1979) 288

\bibitem{pseudo3}
A. Barducci and L. Lusanna, 
{\it J. Phys.} {\bf A16} (1983) 1993

\bibitem{pseudo4}
P. S. Howe, S. Penati, M. Pernici and
P. K. Townsend, {\it Phys. Lett.}
{\bf B215} (1988) 555,
{\it Class. Quantum Grav.} {\bf 6} (1989) 1125

\bibitem{pseudo5}
R. Marnelius and J. E. Martensson,
{\it Nucl. Phys.} {\bf B321} (1989) 185

\bibitem{pseudo6}
M. S. Plyushchay, {\it Mod. Phys. Lett.}
{\bf A8} (1993) 937

\bibitem{pseudo7}
D. M. Gitman, A. E. Gon\c calves and I. V. Tyutin,
{\it Phys. Rev.} {\bf D50} (1994) 5439;\\
G. V. Grigoryan, R. P. Grigoryan and I. V. Tyutin,
{\it Phys. Atom. Nucl.} {\bf 59} (1996) 2212;\\
D. M. Gitman, Preprint hep-th/9608180;\\
D. M. Gitman and I. V. Tyutin,
{\it Int. J. Mod. Phys.} {\bf A12} (1997) 535
  
\bibitem{pr}
M. S. Plyushchay and A. V. Razumov,
{\it Int. J. Mod. Phys.} {\bf A11} (1996) 1427

\bibitem{pt}
R. Jackiw and S. Templeton,
{\it Phys. Rev.} {\bf D23} (1981) 2291

\bibitem{sch}
J. F. Schonfeld, {\it Nucl. Phys.} {\bf B185} (1981)
157

\bibitem{sie}
W. Siegel, {\it Nucl. Phys.} {\bf B156}
(1979) 135

\bibitem{sun}
K. Sundermeyer, {\it Lecture Notes in Physics}
Vol. 169 (Springer, Berlin, 1982)

\bibitem{hht}
J. C. Henty, P. S. Howe and P. K. Townsend,
{\it Class. Quantum Grav.} {\bf 5} (1988) 807 

\bibitem{pre}
Kh. S. Nirov and M. S. Plyushchay,
in preparation.

\end{thebibliography}
\end{document}